\documentclass[fleqn,usenatbib]{mnras}

\usepackage{newtxtext,newtxmath,amsmath}

\usepackage[T1]{fontenc}
\usepackage{orcidlink}

\DeclareRobustCommand{\VAN}[3]{#2}
\let\VANthebibliography\thebibliography
\def\thebibliography{\DeclareRobustCommand{\VAN}[3]{##3}\VANthebibliography}

\usepackage{graphicx}	
\usepackage{amsmath}	

\title[Erratum to STARFORGE]{Second correction to: STARFORGE: Towards a comprehensive numerical model of star cluster formation and feedback}

\author[Grudi\'{c} et al.]{
Michael Y. Grudi\'{c}\orcidlink{0000-0002-1655-5604}$^{1}$\thanks{mgrudic@flatironinstitute.org},
D\'avid Guszejnov\orcidlink{0000-0001-5541-3150}$^{2}$,
Philip F. Hopkins\orcidlink{0000-0003-3729-1684}$^{3}$,
Stella S. R. Offner\orcidlink{0000-0003-1252-9916}$^{4}$, and\newauthor
Claude-Andr{\'e} Faucher-Gigu{\`e}re\orcidlink{0000-0002-4900-6628}$^{5}$
\\
$^{1}$Center for Computational Astrophysics, Flatiron Institute, 162 Fifth Avenue, New York, NY 10010, USA\\
$^{2}$Ab Initio Software, 201 Spring St, Lexington, MA 02421, USA\\
$^{3}$TAPIR, Mailcode 350-17, California Institute of Technology, Pasadena, CA 91125, USA \\
$^{4}$Department of Astronomy, The University of Texas at Austin, TX 78712, USA \\
$^{5}${CIERA and Department of Physics and Astronomy, Northwestern University, 1800 Sherman Ave, Evanston, IL 60201, USA}\\
}

\pubyear{\the\year{}}

\begin{document}
\label{firstpage}
\pagerange{\pageref{firstpage}--\pageref{lastpage}}
\maketitle





\section{The error}

\begin{figure}
\includegraphics[width=0.5\textwidth]{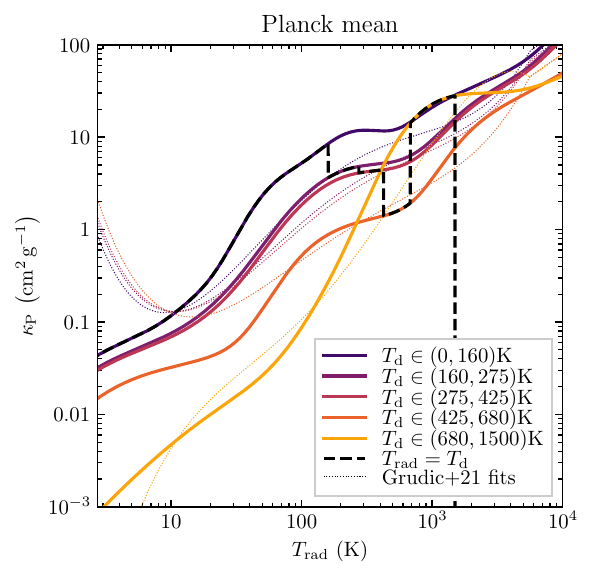}
\vspace{-8mm}
\caption{Planck-mean opacity as a function of both dust temperature $T_{\rm d}$ and radiation temperature $T_{\rm rad}$, computed from the monochromatic opacity tables of \citet{semenov_2003} for their `porous 5-layered sphere' model. The dotted lines in corresponding colours plot the fit given in \citet{starforge_methods}. The dashed line plots the Planck-mean opacity assuming $T_{\rm d} = T_{\rm rad}$, which disagrees with $\kappa_{\rm P}\left(T_{\rm d},T_{\rm rad}\right)$ in general. }
\label{fig:planck}
\end{figure}

The original {\small STARFORGE} simulations methods paper \citep{starforge_methods} provided a fitting function for the infrared Planck-mean dust opacity $\kappa_{\rm P}$ as a function of dust temperature $T_{\rm d}$ and radiation temperature $T_{\rm rad}$ in Appendix C, derived from the model of \citet{semenov_2003}. The fit was performed as a piecewise fourth-order log-log polynomial fit, but fit over a limited range of $T_{\rm rad}$. This fitting function diverges unphysically for physically-reasonable values of $T_{\rm rad} < 10 \rm K$ (Fig. \ref{fig:planck}) and thus is not recommended for general use. Subsequent simulations used an ad-hoc limiter to cap the opacity to a $\propto T_{\rm rad}^2$ power-law at low $T_{\rm rad}$, but we have since updated to a more-accurate table interpolant based upon the calculations presented here.


While revisiting this issue, we found several instances where the distinction between $T_{\rm rad}$ and $T_{\rm d}$ in setting the dust opacity out of local thermodynamic equilibrium (LTE) was neglected \citep{dopcke_2011, grackle, kannan_2020_rt, rigel,zimmermann_2025}, sometimes not even treating the explicit $T_{\rm rad}$ dependence at all. Notably, this nuance was previously pointed out in a footnote in \citet{Cunningham_2018_feedback}, although they did not track the distinct temperatures in their simulation. The issue is subtle, but important to get right because the temperature structure of dust affects the Jeans mass at densities relevant for fragmentation ($n_{\rm H} \gtrsim 10^5 \,\rm cm^{-3}$).

We take this opportunity to clarify the use of Planck-mean opacities in the calculation of dust radiative processes out of LTE, and explicitly compute the dependence of mean opacity on $T_{\rm rad}$ and $T_{\rm d}$.

\section{Planck-mean dust opacities for emission and absorption}

Assuming the heat capacity of dust is very small, the steady-state dust energy equation balances three processes:

\begin{equation}
    \underbrace{\int  \mathrm{d} \nu\, \kappa_\nu \left(T_{\rm d}\right) \rho c u_{\nu}}_{\text{Absorption} }
    - \underbrace{\int  \mathrm{d} \nu\, \epsilon_\nu }_{\text{Emission} }
    + \underbrace{n_{\rm H}^2\alpha_{\rm gd}\left(T\right) \left(T - T_{\rm d}\right)}_{\text{Gas-dust collisions}} = 0,
    \label{eq:dustenergy}
\end{equation}
where $\nu$ is the photon frequency, $\kappa_{\nu}\left(T_{\rm d}\right)$ is the monochromatic dust absorption opacity, $\rho$ is the mass density, $u_\nu$ is the photon energy density per unit frequency, $\epsilon_\nu$ is the volumetric emissivity, $n_{\rm H}$ is the number density of H nuclei, $T$ is the gas temperature, and $\alpha_{\rm gd}$ is the gas-dust collision coefficient \citep{hollenbach_mckee_1989}. Here we have made explicit the dependence of $\kappa_{\nu}\left(T_{\rm d}\right)$ upon dust temperature, due to varying grain composition as volatiles sublimate \citep{semenov_2003}.

The assumption we make for the {\small STARFORGE} far-IR photon frequency component is 
\begin{equation}
u_\nu = u_{\rm IR} \times \frac{B_\nu\left(T_{\rm rad}\right)}{\int \mathrm{d} \nu\, B_\nu\left(T_{\rm rad}\right)}
\end{equation}
 i.e. the photon energy distribution is proportional to that of a black-body with temperature $T_{\rm rad}$, with frequency-integrated energy density $u$. Integrating Eq. \ref{eq:dustenergy} while neglecting the other radiation bands absorbed by dust:
\begin{equation}
     \kappa_P\left(\underline{T_{\rm d}},\underline{T_{\rm rad}}\right) \rho c u_{\rm IR}
    - \epsilon_{\rm d}
    + n_{\rm H}^2\alpha_{\rm gd}\left(T\right) \left(T - T_{\rm d}\right) = 0,
\end{equation}
where 
\begin{equation}
\kappa_{\rm P}\left(\underline{T_{\rm d}},\underline{T_{\rm rad}}\right) = \frac{\int \mathrm{d} \nu\, \kappa_\nu 
\left(T_{\rm d}\right) B_\nu\left(T_{\rm rad}\right)}{\int \mathrm{d} \nu\, B_{\nu}\left(T_{\rm rad}\right)}
\end{equation}
is the Planck-mean opacity, which has two distinct temperature arguments: the first accounts for variations in dust composition with $T_{\rm d}$, and the second dependence upon $T_{\rm rad}$ due to the original frequency-dependence of $\kappa_\nu\left(T_{\rm d}\right)$.

In local thermodynamic equilibrium where $T_{\rm d} = T_{\rm rad} = T$ and $u=a T^4$, we require $\epsilon_{\rm d} = a c \rho T^4 \kappa_{\rm P}\left(T,T\right)$. In general, out of of LTE, Kirchoff's law for thermal emission therefore implies that $\epsilon_{\rm d} = a c \rho \kappa_{\rm P}\left(T_{\rm d},T_{\rm d}\right) T_{\rm d}^4$. So the frequency-integrated energy balance equation becomes

\begin{equation}
     \rho c \left(\kappa_{\rm P}\left(T_{\rm d},T_{\rm rad}\right) u_{\rm IR} - \kappa_P\left(T_{\rm d},T_{\rm d}\right) a T_{\rm d}^4\right)  + n_{\rm H}^2\alpha_{\rm gd}\left(T\right) \left(T - T_{\rm d}\right)=0.
\label{eq:dustenergy2}
\end{equation}
Thus, the same functional form for $\kappa_{\rm P}$ is used for both emission and absorption, but the emission term substitutes $T_{\rm d}$ in the $T_{\rm rad}$ slot while the absorption term uses the distinct temperatures. It is apparent that $T_{\rm rad} \sim T_{\rm dust}$ only under certain conditions, e.g. when the radiative terms are dominant and $u_{\rm IR} \approx a T_{\rm rad}^4$. An important counterexample is at high attenuations deep within a pre-stellar molecular cloud, where the ambient optical and UV components are attenuated. Here $u_{\rm IR}$ is dominated by the FIR dust-emission component of the ISM, which is highly diluted compared to a blackbody and hence $T_{\rm d} << T_{\rm rad}$. This regime of dust energy balance is important for thermal evolution during pre-stellar core collapse \citep{masunaga1998,hennebelle_imf_review}.

For completeness, the full equation solved in the current version of the {\small STARFORGE} model for $T_{\rm d}$, accounting for all frequency components and radiative processes, is 
\begin{equation}
\begin{split}
 & \rho c \left[\sum_i \kappa_i u_i + \kappa_{\rm P}\left(T_{\rm d},T_{\rm rad}\right)  u_{\rm IR} - \kappa_{\rm P}\left(T_{\rm d},T_{\rm d}\right) a T_{\rm d}^4\right]  \\
 &+ n_{\rm H}^2\alpha_{\rm gd}\left(T\right) \left(T - T_{\rm d}\right)=0,
 \end{split}
\end{equation}
where $i$ runs over the FUV, near-UV, and optical-NIR frequency bands, with corresponding dust opacities $\kappa_i$.

\section{Mean dust opacity as a function of $T_{\rm rad}$ and $T_{\rm d}$}

We compute the Planck-mean opacity explicitly as a function of $T_{\rm rad}$ and $T_{\rm d}$ by numerically integrating the monochromatic opacity tables provided by \citet{semenov_2003} for all 5 $T_{\rm d}$ regions demarcated by sublimation points of the dust components. This routine is implemented in the \texttt{radiation.dust} submodule of the \texttt{meshoid} Python package\footnote{\url{https://github.com/mikegrudic/meshoid}}. However, it is not practical to compute $\kappa_{\rm P}$ in this way on-the-fly within the dust temperature solver in {\small GIZMO} or another RHD simulation. Instead, we use a simple log-space linear interpolant of a lookup table as a function of $T_{\rm rad}$, with a separate table for each of the 5 distinct temperature regions. 

\begin{figure}
\includegraphics[width=0.5\textwidth]{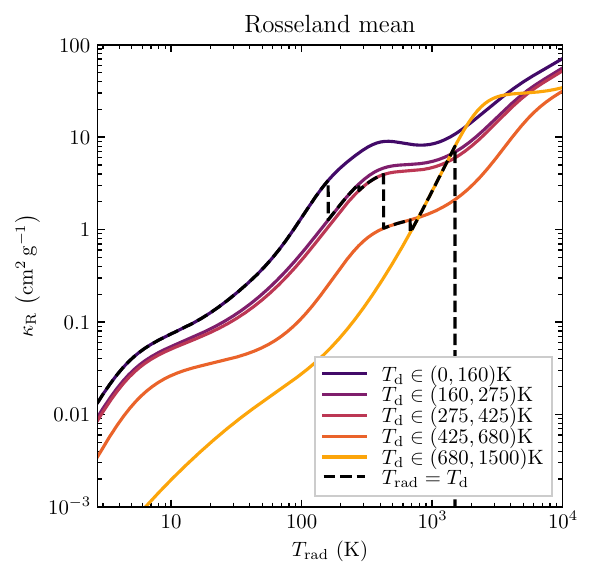}
\vspace{-8mm}
\caption{Rosseland mean dust opacity as a function of both dust temperature $T_{\rm d}$ and radiation temperature $T_{\rm rad}$, computed from the opacity tables of \citet{semenov_2003} for their `porous 5-layered sphere' model.}
\label{fig:rosseland}
\end{figure}

For completeness, we also also compute Rosseland-mean opacities (Fig. \ref{fig:rosseland}) as a function $T_{\rm rad}$ and $T_{\rm d}$. In the radiative diffusion regime where the Rosseland mean is useful, conditions are likely to be closer to LTE and distinction between $T_{\rm rad}$, $T$, and $T_{\rm dust}$ is typically less important.

The code used to generate these figures and a set of tabulated mean opacities are available at \url{https://github.com/mikegrudic/STARFORGE-methods-errata} or \url{https://doi.org/10.5281/zenodo.17596225}.

\bibliographystyle{mnras}
\bibliography{bibliography} 


\bsp	
\label{lastpage}
\end{document}